\documentclass[a4paper,aps,prd,showpacs,nofootinbib,preprintnumbers,amsmath,amssymb,mcite,superscriptaddress,12pt]{revtex4-1}

\usepackage{comment}

\usepackage[table]{xcolor}
\usepackage{graphicx}% Include figure files
\usepackage{subfigure}
\usepackage{dcolumn}% Align table columns on decimal point
\usepackage{bm}% bold math
\usepackage{esvect}
\usepackage{accents}
\usepackage{slashed}

\usepackage{hyperref}
\hypersetup{colorlinks=true,linkcolor=magenta,anchorcolor=green,citecolor=cyan,filecolor=black,menucolor=black,urlcolor=brown}
\usepackage{slashed}
\newcommand{\be}{\begin{equation}}
\newcommand{\ee}{\end{equation}}
\newcommand{\ba}{\begin{eqnarray}}
\newcommand{\ea}{\end{eqnarray}}

\newcommand{\Mpl}{M_{\textrm{Pl}}}

\usepackage{braket}

\begin{document}

\title{Hide and Seek: Screened Scalar Fields in Hydrogen and Muonium}

\author{Philippe Brax}
\affiliation{Institut de Physique Th\'eorique, Universit\'e  Paris-Saclay, CEA, CNRS, F-91191 Gif-sur-Yvette Cedex, France}
\author{Anne-Christine Davis}
\affiliation{DAMTP, Centre for Mathematical Sciences, University of Cambridge, Wilberforce Road, Cambridge CB3 0WA, United Kingdom}
\author{Benjamin Elder}
\affiliation{Department of Physics and Astronomy, University of Hawai'i, 2505 Correa Road, Honolulu, HI 96822, USA}
\begin{abstract} \footnotesize
We compute bounds on screened scalar field theories from hydrogen-like systems.  New light scalar fields generically have a direct coupling to matter.  Such a coupling is strongly constrained by myriad experimental measurements.  However, certain theories possess a {\it screening mechanism} that allows the effects of this coupling to weaken dynamically, and to evade many such bounds.  We compute the perturbations to the energy levels of hydrogen-like systems due to screened scalar fields.  We then use this result in two ways.  First, we compute bounds from hydrogen spectroscopy, finding significantly weaker bounds than have been  reported before as screening effects were overlooked.  Second, we show that muonium is an intrinsically much more sensitive probe of screened scalar fields.
For chameleon models, muonium experiments probe a large part of the parameter space that is as yet unexplored by low energy physics and has so far only been tested by high-energy particle physics experiments.
\end{abstract}

\maketitle

\section{Introduction}
Einstein's theory of gravity, with a small cosmological constant, and the Standard Model of particle physics are both staggeringly successful models of our universe.  Nevertheless, it is still useful to consider additions to these models, as there are multiple well-known puzzles associated with the standard paradigm.  For instance, the accelerated expansion of the universe has motivated a wide range of models that modify our theory of gravity (for recent reviews, see ~\cite{Clifton:2011jh, Bull:2015stt, CANTATA:2021ktz}).  Likewise, new physics has also recently been invoked to explain the 4.2-$\sigma$ tension between the Standard Model prediction~\cite{Aoyama:2012wk,Aoyama:2019ryr,Czarnecki:2002nt,Gnendiger:2013pva,Davier:2017zfy,Keshavarzi:2018mgv,Colangelo:2018mtw,Hoferichter:2019mqg,Davier:2019can,Keshavarzi:2019abf,Kurz:2014wya,Melnikov:2003xd,Masjuan:2017tvw,Colangelo:2017fiz,Hoferichter:2018kwz,Gerardin:2019vio,Bijnens:2019ghy,Colangelo:2019uex,Blum:2019ugy,Colangelo:2014qya} and experimental measurements~\cite{Muong-2:2006rrc,Muong-2:2021ojo} of the muon's anomalous magnetic dipole moment, the observed electron recoil in XENON1T~\cite{XENON:2020rca}, and the planar arrangement of satellite galaxies~\cite{Naik:2022lcn}. These puzzles justify the addition of at least one new particle species to alleviate tension between theory and experiment.

The minimal addition is a single degree of freedom, typically in the form a new scalar field.  In order for this approach to work it is important that this addition does not create new tensions.  Crucially, a generic feature of new scalar fields is an explicit coupling to matter.   This coupling enables the new scalar to mediate a ``fifth force'' between matter particles, which is subject to strong constraints from laboratory, solar system, and astrophysical tests of gravity.  The conventional wisdom is that the new scalar's coupling to matter fields must be exceedingly weak, or the scalar particle's mass must be sufficiently large that the fifth force is too short-ranged to be detectable.

A powerful loophole to this argument exists, however.  For certain matter couplings and scalar self-interactions, the fifth force can weaken dynamically.  This property, dubbed a ``screening mechanism'', enables a theory with intrinsically strong, and long-ranged, fifth forces to pass experimental tests~\cite{Joyce:2014kja}.  There are three broad classes of screening mechanisms known to date.  First, there is chameleon screening~\cite{Khoury:2003aq}, in which scalar particles' mass becomes large in dense environments, making the fifth force short-ranged.  Second, symmetron screening~\cite{Damour:1994zq, Olive:2007aj, Hinterbichler:2010es} involves a matter coupling that depends on the average ambient matter density, such that the field decouples in dense environments.  Finally, there is derivative-induced  screening, for instance the Vainshtein mechanism~\cite{Vainshtein:1972sx,Babichev:2013usa} of massive gravity~\cite{Hinterbichler:2011tt,deRham:2014zqa} and galileons~\cite{Nicolis:2008in}, for example, in which the fifth force law departs from $1/r^2$ at near distances.

Since their proposal approximately fifteen years ago, a number of laboratory experiments and astrophysical observations have targeted screened theories and have made great progress in constraining their available parameter space.  Some of most relevant laboratory tests include atom interferometers~\cite{Burrage:2014oza,Hamilton:2015zga,Elder:2016yxm,Burrage:2016rkv, Jaffe:2016fsh, Sabulsky:2018jma}, cold bouncing neutrons~\cite{Cronenberg:2018qxf,Brax:2011hb,Brax:2013cfa}, electron $g - 2$ measurements~\cite{Brax:2018zfb}, and torsion balances~\cite{Upadhye:2012qu,Upadhye:2012rc}, all of which have been recently reviewed in \cite{Burrage:2016bwy,Burrage:2017qrf,CANTATA:2021ktz,Brax:2021wcv}.  Although these theories were originally motivated by dark energy, they have since found application in other arenas, such as the effects of dark matter~\cite{Burrage:2016yjm,Salzano:2016udu,OHare:2018ayv,Burrage:2018zuj,Chen:2019kcu}, or to explain the muon's anomalous magnetic moment~\cite{Brax:2021owd} and the electron recoil in XENON1T~\cite{Vagnozzi:2021quy}.

One of the most stringent bounds to date on new scalar fields has derived from hydrogen spectroscopy.  If the scalar coupling to matter is strong enough, and the scalar particle's Compton wavelength is longer than 1 {\AA}, then the new scalar field behaves as a potential that perturbs the energy levels in the hydrogen atom.  This was computed for a generic scalar field in \cite{Brax:2010gp}, implicitly assuming the hydrogen nucleus was unscreened.  However, this bound does not extend in a straightforward manner to screened theories, as has been sometimes claimed.  This bound essentially treats the hydrogen nucleus as a pointlike particle that is therefore exempt from screening. The resulting bounds are too stringent; an appropriate treatment would account for the finite size and radius of the hydrogen nucleus and the resulting screening effect of the fifth force sourced by it.

The purpose of this paper is twofold.  First, we revisit the spectroscopy bounds of \cite{Brax:2010gp} within the context of chameleon and symmetron theories, carefully accounting for the screening behavior of the scalar field around the nucleus.  We will find that this process relaxes the bounds considerably.  Second, we will consider a closely related system that turns out to be a far more sensitive probe of screened theories: muonium.  This system, composed of an antimuon and an electron, has a spectrum nearly identical to that of hydrogen, yet is composed entirely of fundamental particles.  As the constituent particles are truly pointlike in nature, they are not subject to the screening effect that is found for extended objects like hydrogen nuclei.  We find bounds on screened scalar fields that are limited only by the finite range of the fifth force.

The plan of the paper is as follows.  In Section 2 we derive expressions for a scalar field with a screening mechanism around an extended object.  Our expressions apply only to canonical scalar fields like chameleon or symmetron theories.  We do not consider Vainshtein-screened theories, as a proper treatment would require accounting for the presence of the Earth and Sun nearby, and is therefore outside the scope of this article.  In Section 3 we compute the fifth force perturbations to a hydrogen-like atom's energy levels.  In Sections IV and V we apply our results to compute bounds on chameleon theories from the hydrogen and muonium spectroscopy, respectively.   In Section VI we repeat this analysis for symmetron theories, and we make concluding remarks in Section VII.

{\bf Conventions:} We adopt the mostly-plus metric signature $\eta_{\mu \nu} = \mathrm{diag}(-1, 1, 1, 1)$, define the reduced Planck mass as $\Mpl^2 = (8 \pi G)^{-1}$, and use natural units such that $c = \hbar = 1$.

\section{Screened modified gravity}
\label{sec:review}
In this Section we derive the screened scalar field profile around a spherical source, which in this paper will represent the nucleus of a hydrogen-like system.  This treatment is standard, so the reader already familiar with this calculation may wish to skip ahead to the final result of Eq.~\eqref{field-profile}.

We begin with the action for a canonical scalar field with an explicit coupling to the local density $\rho$ of matter fields which are non-relativistic:
\begin{equation}
    {\cal L} = - \frac{1}{2} (\partial \phi)^2 - V(\phi) - A(\phi) \rho~.
\end{equation}
The field dynamics respond to the ``effective potential''
\begin{equation}
    V_\mathrm{eff}(\phi) = V(\phi) + A(\phi) \rho~.
\end{equation}
The scalar field equation of motion is
\begin{equation}
    \Box \phi = V_{\mathrm{eff},\phi}(\phi)~,
\end{equation}
where the comma denotes differentition: $f,_\phi \equiv \frac{\mathrm{d}}{\mathrm{d}\phi} f$~.
The scalar field mediates an attractive force between objects, causing a test particle to accelerate as
\begin{equation}
    \vec a = - \vec \nabla \Phi_\mathrm{N} - \vec \nabla A(\phi)~,
    \label{geodesic}
\end{equation}
where $\Phi_\mathrm{N}$ is the Newtonian gravitational potential.\footnote{Strictly speaking the fifth force is given by $-\vec \nabla \ln A$ which coincides with $-\vec \nabla A$ when $A$ is close to unity as required to avoid large deviations from Newton's law.}
As such, the coupling function $A(\phi)$ behaves as a potential for the fifth force.

Let us solve for the field around a spherical source mass, surrounded by a uniform density background:
\begin{equation}
    \rho = \begin{cases}
    \rho_\mathrm{in} & r \leq R \\
    \rho_\mathrm{out} & r > R~.
    \end{cases}
\end{equation}
We solve in a piecewise manner, inside and outside the source.  First we expand around a fixed value $\bar \phi$:
\begin{equation}
    \phi = \bar \phi + \varphi~.
\end{equation}
The value we choose for $\bar \phi$ allows us to handle the several different cases.

%\subsection{Expand around a potential minimum}
If $\bar \phi$ corresponds to a minimum of $V_\mathrm{eff}(\phi)$, then the leading terms in the expanded action are
\begin{equation}
    {\cal L} = - \frac{1}{2} (\partial \varphi)^2 - \frac{1}{2} V_{\mathrm{eff}, \phi\phi}\bigg|_{\phi = \bar \phi} \varphi^2~.
\end{equation}
The quadratic term behaves as a mass for the perturbations $\varphi$, hence we define the ``effective mass'' as
\begin{equation}
    m^2_\mathrm{eff}(\phi) \equiv V_{\mathrm{eff}, \phi \phi}(\phi)~,
\end{equation}
in terms of which the equation of motion is
\begin{equation}
    \Box \varphi = m_\mathrm{eff}^2 \varphi~.
\end{equation}
%\subsection{Expand about non-equilibrium point}
On the other hand, if $\bar \phi$ does not minimise $V_\mathrm{eff}$, then the leading terms in the expansion are
\begin{equation}
    {\cal L} = - \frac{1}{2} (\partial \varphi)^2 - V_{\mathrm{eff}, \phi}(\bar \phi) \varphi~.
\end{equation}
This behaves like a source for the perturbations $\varphi$, hence we define the term ``effective source'' as\footnote{ The effective source has mass dimension three here. It usually coincides with $\rho/M$ where $\rho$ is the matter density of the source, either $\rho_{\rm in}$ or $\rho_{\rm out}$, and $M$ is a coupling scale.}
\begin{equation}
    \rho_\mathrm{eff}(\phi) = V_{\mathrm{eff}, \phi}(\phi)~.
\end{equation}

\subsection{Case 1: field becomes heavy inside the source}
In the first case we consider, we expand about $\phi_\mathrm{in}$ inside the sphere and $\phi_\mathrm{out}$ outside the sphere, where both field values minimise the effective potentials in their respective regions.  We also define the effective masses about those minima as $m_\mathrm{in, out}$.  Then the solution is
\begin{equation}
    \phi = \begin{cases}
    \phi_\mathrm{in} + \tilde{A} \frac{\sinh{m_\mathrm{in} r}}{r} & r \leq R~, \\
    \phi_\mathrm{out} + \tilde B\frac{e^{- m_\mathrm{out} r}}{r} & r > R~.
    \end{cases}
\end{equation}
We can solve for the integration constants $\tilde{A}, \tilde B$ by matching $\phi$ and its derivative at $r = R$.  However, we are only interested in the fifth force outside the object, so we only need $\tilde B$:
\begin{equation}
    \tilde B = - (\phi_\mathrm{out} - \phi_\mathrm{in}) R e^{m_\mathrm{out} R} \left( \frac{m_\mathrm{in} R - \tanh m_\mathrm{in} R}{m_\mathrm{in} R + m_\mathrm{out} R \tanh m_\mathrm{in} R} \right)~.
\end{equation}
This result is not particularly edifying, so let us make some simplifying assumptions.  First, in cases of screening, we typically have $|\phi_\mathrm{in}| \ll |\phi_\mathrm{out}|$, otherwise there would not be much fifth force to speak of.  Furthermore, we have in mind $m_\mathrm{out} R \lesssim 1$, without which the fifth force would be very short-ranged relative to the size of the system being considered.  These two assumptions leave us with
\begin{equation}
    \tilde B = - \phi_\mathrm{out} R \left(1 - \frac{\tanh m_\mathrm{in} R}{m_\mathrm{in} R} \right)~.
\end{equation}
Let us now take the limit in which the field is very heavy inside the source.  That is, the field's Compton wavelength inside the object is is much shorter than the object's radius, $m_\mathrm{in} R \gg 1$, and consequently the field easily reaches $\phi_\mathrm{min}$.  In this case, we have
\begin{equation}
    \phi(r > R) = \phi_\mathrm{out} - \phi_\mathrm{out} R \frac{e^{- m_\mathrm{out} r}}{r}~.
\end{equation}
This leaves us with the quintessential screened field profile of a Yukawa field profile outside a spherical object, subject to the boundary condition that $\phi \approx 0$ at the surface of the sphere.

\subsection{Case II: Field remains light everywhere}
If the field remains light inside the source, $m_\mathrm{in} R \ll 1$, then the field cannot deviate far from $\phi_\mathrm{out}$ even deep inside the source.  In this case, it is not sensible to expand about $\phi_\mathrm{in}$ inside the source, as was done in the previous subsection.  Instead, we will expand about $\phi_\mathrm{out}$ both inside and outside the source.

Inside the source, then, we expand about $\phi_\mathrm{out}$, which is {\it not} the minimum of $V_\mathrm{eff}(\phi)$ in this region.  Consequently, the equation of motion for perturbations inside the source takes the form
\begin{equation}
    \Box \varphi = V,_\phi(\phi_\mathrm{out}) + A,_\phi(\phi_\mathrm{out}) \rho_\mathrm{in}~.
    \label{eom-in-intermediate}
\end{equation}
The right hand side is further simplified in the following way.  Far away from the source, the field is in equilibrium, so
\begin{equation}
    |A,_\phi(\phi_\mathrm{out}) \rho_\mathrm{out}| = |V,_\phi(\phi_\mathrm{out})|~.
\end{equation}
Inside the source, we have
\begin{equation}
    |A,_\phi(\phi_\mathrm{out}) \rho_\mathrm{in}| \gg |A,_\phi(\phi_\mathrm{out}) \rho_\mathrm{out}| = |V,_\phi(\phi_\mathrm{out})|~,
\end{equation}
where we have used the fact that the source is much denser than its surroundings, $\rho_\mathrm{in} \gg \rho_\mathrm{out}$.  This justifies dropping the first term on the right hand side of Eq.~\eqref{eom-in-intermediate}, and the solution is
\begin{equation}
    \phi = \begin{cases}
    \phi_\mathrm{out} + A,_\phi(\phi_\mathrm{out}) \rho_\mathrm{in} \frac{r^2}{6} + \tilde{A} & r \leq R~, \\
    \phi_\mathrm{out} + \tilde{B}\frac{e^{- m_\mathrm{out} r}}{r} & r > R~,
    \end{cases}
\end{equation}
where $\tilde A, \tilde B$ are integration constants.  These are fixed by matching the field and its first derivative at $r = R$, leaving us with the exterior solution
\begin{equation}
    \phi(r > R) = \phi_\mathrm{out} - A,_\phi(\phi_\mathrm{out}) \frac{m_\mathrm{src}}{4 \pi} \frac{e^{-m_\mathrm{out} r}}{r}~.
\end{equation}

We have now solved the scalar field in two limiting cases: first, where the field is heavy inside the source and second, where the field remains very light inside the source.  It is convenient to encompass both of these limiting cases with the single expression
\begin{align} \nonumber
    \phi(r > R) &= \phi_\mathrm{out} - A,_\phi(\phi_\mathrm{out}) \lambda_\mathrm{src} m_\mathrm{src} \frac{e^{- m_\mathrm{out} r}}{4 \pi r}~, \\
    \lambda_\mathrm{src} &\equiv \min \left( \frac{3 \phi_\mathrm{out}}{A,_\phi(\phi_\mathrm{out}) \rho_\mathrm{in} R^2}, 1 \right)~,
    \label{field-profile}
\end{align}
where we have introduced the ``screening factor'' of the source $\lambda_\mathrm{src}$.  The total scalar charge carried by the object is $Q_\mathrm{src} = \lambda_\mathrm{src} m_\mathrm{src}$.

It is seen that the screening behavior and coupling strength depend on the ambient scalar field value $\phi_\mathrm{out}$.  Determining $\phi_\mathrm{out}$ for a specific setup, such as a vacuum chamber, requires accounting for the non-linear dynamics of the theory, and is consequently a model-dependent quantity.  We will discuss how to determine $\phi_\mathrm{out}$ in specific examples as we encounter them.

\section{Perturbed energy levels}

In this section we compute the perturbation to a hydrogenic atom's energy levels due to the new scalar fifth force.  If the fifth force contribution to the electron's Hamiltonian is $\delta H$, the perturbation to the electron's energy levels are given by the classic formula
\begin{equation}
    \delta E_n = \braket{\psi_n | \widehat{\delta H} | \psi_n}~,
    \label{energy-perturbation}
\end{equation}
where $\psi_n$ is the wavefunction of the $n^\mathrm{th}$ energy level.  We will only be interested in the first and second energy levels, for which we have:
\begin{align} \nonumber
    \psi_{1s} &= \frac{1}{\sqrt \pi} \left( \frac{Z}{a_0} \right)^{3/2} e^{- Z r / a_0}~, \\
    \psi_{2s} &= \frac{1}{4 \sqrt{2 \pi}} \left( \frac{Z}{a_0} \right)^{3/2} \left(2 - \frac{Z r}{a_0} \right) e^{- Z r / a_0}~,
\end{align}
where $a_0$ is the Bohr radius and $Z = 1$ is the atomic number of the hydrogenic atom.

The perturbation to the electron's Hamiltonian\footnote{In this work we treat the electron as completely non-relativistic, as is appropriate for hydrogen-like systems.  A relativistic treatment would start with the Dirac Lagrangian for electron~\cite{Brax:2010jk,Wong:2017jer}.} follows from Eq.~\eqref{geodesic} as
\begin{equation}
    \delta H = m_e A(\phi)~.
\end{equation}
Taylor expanding this expression to linear order about the ambient field value $\phi_\mathrm{out}$ and dropping the irrelevant constant term yields
\begin{align} \nonumber
    \delta H &=  m_e A,_\phi(\phi_\mathrm{out}) \left( \phi - \phi_\mathrm{out} \right) \\
    &=- A,_\phi(\phi_\mathrm{out})^2 \lambda_N m_N m_e \frac{e^{-m_\mathrm{out} r}}{4 \pi r}~,
\end{align}
where in the second line we have used the field profile in Eq.~\eqref{field-profile}.  Here $\lambda_N$ and $m_N$ refer to the screening factor and mass of the atom's nucleus, respectively.

The 1s and 2s energy level perturbations may now be computed via Eq.~\eqref{energy-perturbation}, and are 
\begin{align} \nonumber
    \delta E_{1s} &= A,_\phi(\phi_\mathrm{out})^2 \frac{ \lambda_N m_N m_e}{\pi} \frac{Z^3}{a_0^3 \left(\frac{2 Z}{a_0} + m_\mathrm{out} \right)^2}~, \\ 
    \delta E_{2s} &= A,_\phi(\phi_\mathrm{out})^2 \frac{\lambda_N m_N m_e}{16 \pi} \frac{Z^3 \left( Z^2 + 2 a_0^2 m_\mathrm{out}^2 \right)}{a_0 \left(Z + a_0 m_\mathrm{out} \right)^4}~.
    \label{energy-levels}
\end{align}
These perturbations shift the energy gap between the 1s and 2s energy levels by an amount
\begin{equation}
    \delta E_{1s-2s} = |\delta E_{2s} - \delta E_{1s} |~.
    \label{energy-difference}
\end{equation}
Taking the unscreened limit in which $\lambda_N \to 1$ and $m_\mathrm{out} r \ll 1$, we obtain for hydrogen ($Z = 1$)
\begin{equation}
    \delta E_{1s-2s} = A,_\phi(\phi_\mathrm{out})^2 \frac{3 m_N m_e}{16 \pi a_0}~,
    \label{dE-unscreened}
\end{equation}
precisely matching the result of \cite{Brax:2010gp}.  This serves as a useful check, but in what follows we will mostly use the more general expressions in Eqs.~\eqref{energy-levels} and \eqref{energy-difference}.

\section{Hydrogen spectroscopy}
The $1s-2s$ transition of hydrogen has been measured to an accuracy of 5 parts in $10^{11}$ ~\cite{PhysRevLett.65.571}, indicating an absolute uncertainty of $\delta E_{1s-2s} = 5.1 \times 10^{-10}~\mathrm{eV}$.  This measurement, along with Eq.~\eqref{energy-difference}, can constrain the parameters of canonical scalar field theories that couple to matter.  For concreteness, in this section and the next we adopt the chameleon as the prototypical model of screened modified gravity.  The simplest chameleon model is characterized by the following self-interaction potential and matter coupling:
\begin{equation}
    V_\mathrm{cham}(\phi) = \frac{\Lambda^5}{\phi}~, \quad \quad A_\mathrm{cham}(\phi) = \frac{\phi}{M}~.
\end{equation}
In the unscreened limit of this theory, we can apply Eq.~\eqref{dE-unscreened} to obtain a constraint from Hydrogen spectroscopy of
\begin{equation}
    M \gtrsim 10~\mathrm{TeV}~,
\end{equation}
as has been reported previously for unscreened fields~\cite{Brax:2010gp}.

To extend bounds into the regime where screening can occur we will use the more general result of Eq.~\eqref{energy-difference}, which relies on further experimental details.  The proton, of course, has finite radius and density, and can therefore be screened.  Its screening factor must be computed via Eq.~\eqref{field-profile}.

As seen in Section~\ref{sec:review}, screening behavior relies on the ambient scalar field value $\phi_\mathrm{out}$.  For the specific case of a chameleon field, this is determined in the following way.  First, one computes the minimum of the effective potential inside a region of constant density $\rho$:
\begin{equation}
    \phi_\mathrm{min} = \sqrt{ \frac{M \Lambda^5}{\rho} }~.
\end{equation}
The walls of the vacuum chamber are made of dense metal, and the density inside the vacuum chamber is tiny, so in general the field rolls from a small value near the walls to a large value in the center.  If there is sufficient room inside the vacuum chamber, the field levels off at $\phi_\mathrm{min}$ towards the middle of the chamber.  This rolling behavior takes place over a distance $\approx m_\mathrm{out}^{-1}$.  However, if the vacuum chamber is smaller than $m_\mathrm{out}^{-1}$, then there is not sufficient room for this to take place, and instead the field will only reach a central value $\phi_\mathrm{vac}$ such that $R_\mathrm{vac} \approx m_\mathrm{eff}(\phi_\mathrm{vac})^{-1}$~\cite{Burrage:2014oza,Hamilton:2015zga,Elder:2016yxm}.  This corresponds to a field value
\begin{equation}
    \phi_\mathrm{vac} = \xi \left( 2 \Lambda^5 R_\mathrm{vac}^2 \right)^{1/3}~,
\end{equation}
where $\xi$ is an $O(1)$ factor that depends on the geometry of the vacuum chamber.  For a spherical vacuum chamber, $\xi = 0.55$.

To summarise, the field rolls from a small value near the vacuum chamber walls to $\phi_\mathrm{min}$ or $\phi_\mathrm{vac}$ in the center, whichever is smaller, so we have
\begin{equation}
    \phi_\mathrm{out} = \min \left( \phi_\mathrm{vac}, \phi_\mathrm{min} \right)~.
\end{equation}
Alternatively, one could compute the field value $\phi_\mathrm{out}$ numerically, as was done in \cite{Elder:2016yxm}.

\begin{figure}[t]
    \centering
    \includegraphics[width=0.75\textwidth]{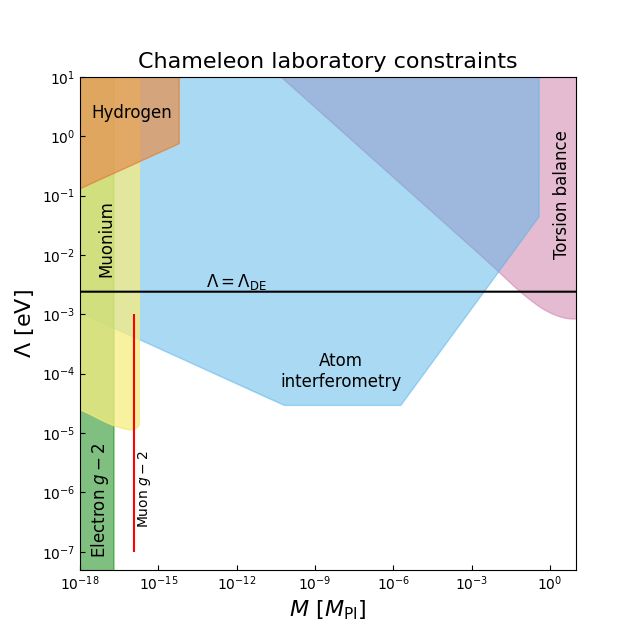}
    \caption{\scriptsize Bounds on chameleon theory parameter space.  Before this work, the right edge of the hydrogen bound extended indefinitely downwards.  We have found that accounting for the screening of the proton results in significantly weakened hydrogen constraints.  Conversely, we find that muonium bounds are limited only by the Compton wavelength of the chameleon particle.  The red line indicates chameleon models that alleviate the muon $g - 2$ tension \cite{Brax:2021owd}.   This line extends vertically downwards to $\Lambda \approx 10^{-7}~\mathrm{eV}$.  We see that muonium rules out approximately half of those models.}
    \label{fig:chameleon-bounds}
\end{figure}

With the details of how to compute $\phi_\mathrm{out}$ in hand, we can use Eqs.~\eqref{energy-levels} and \eqref{energy-difference} along with the experimental values in Table~\ref{tab:params} to compute the constrained region of parameter space.  This was done numerically, resulting in the bound shown in Fig.~\ref{fig:chameleon-bounds}.  We see the vertical boundary corresponding to the unscreened limit of the theory, as well as a slanted line which is a result of the screening of the proton.

\begin{table}[]
    \centering
    {\footnotesize
    \begin{tabular}{| l || c | c | c | c |}
        \hline
        Quantity & Symbol & Hydrogen & Muonium & Units \\
        \hline
         Uncertainty & $\delta E_{1s-2s}$ & $5.1 \times 10^{-10}$ & $4.1 \times 10^{-8}$ & eV \\
         Source radius & $R$ & 0.877 & - & fm \\
         Source mass & $m_\mathrm{N}$ & 938 & 106 &  MeV \\
         Vacuum chamber radius  & $R_\mathrm{vac}$  &  25 & 0.8 & cm \\
         Vacuum chamber density  & $\rho_\mathrm{vac}$  &  $6.1 \times 10^{-13}$ & $10^{-13}$ & g/cm$^3$ \\
         \hline
    \end{tabular}
    \caption{\scriptsize Experimental parameters for the measurement of the $1s-2s$ transition energy in hydrogen~\cite{PhysRevLett.65.571} and muonium~\cite{Meyer:1999cx}.  For hydrogen, the density of the gas in the vacuum chamber follows from the reported pressure of $5 \times 10^{-6} ~\mathrm{mbar}$ of hydrogen gas at a temperature of $200~\mathrm{K}$.  For muonium, the density of the gas in the vacuum chamber follows from the reported pressure of $10^{-6} ~\mathrm{mbar}$ of helium gas at a temperature of $296~\mathrm{K}$.  The total size of the vacuum chamber is not reported, but the experiment was performed $8~\mathrm{mm}$ above a surface, so that is taken as the chamber radius.}
    \label{tab:params}
    }
\end{table}

\begin{comment}
\begin{itemize}
    \item Hydrogen measurement Zimmermann 1990 \cite{PhysRevLett.65.571}
    \item Relative uncertainty on the transition $= 5 \times 10^{-11}$
    \item Absolute uncertainty $= 5.1 \times 10^{-10}~\mathrm{eV}$
    \item Vacuum chamber radius $= 25~\mathrm{cm}$.
    \item Density $= 6\times10^{-13}~\mathrm{g/cm}^3$.
    \item Vertical boundary at $M = 6.1 \times 10^{-15}\Mpl$
\end{itemize}
\end{comment}

\section{Muonium spectroscopy}
Muonium is a hydrogen-like system consisting of a single electron orbiting an anti-muon.  The anti-muon has a positive charge and is much heavier than the electron, resulting in an energy spectrum that is nearly identical to that of hydrogen.\footnote{One might wonder about positronium spectroscopy as well, a system composed of an electron and an anti-electron.
%In this case, an electron and anti-electron pair orbit their center of mass.
The experimental uncertainty on the 1s-2s transition energy is approximately 1 order of magnitude smaller than that of muonium, but the source (an electron or a positron) has a mass that is 1 order of magnitude smaller as well, so the bounds on modified gravity are comparable to those from muonium.  As such, we focus on muonium in this paper, as the mechanics of the system are slightly simpler and are nearly identical to hydrogen.}

The key advantage that muonium presents relative to hydrogen is that the source mass, the muon, is a fundamental particle with no internal structure.  That is, the muon has zero finite extent and is always unscreened.\footnote{ Coupling a scalar field with a nonlinear potential to point particles raises subtleties regarding the precise strength of the interaction.  These were considered, along with quantum fluctuations, in \cite{Burrage:2021nys}.  In this work we take the simpler view that point particles are completely unscreened.  This is akin to assuming that the muon-electron interaction is dominated by the Feynman diagram in which a single scalar particle is exchanged.}  Muonium thus represents an intrinsically more sensitive probe to screened scalar fields.

That being said, there are two disadvantages that muonium has relative to hydrogen.  First, the experimental uncertainty on the $1s-2s$ transition energy is larger by a factor of 100.  Second, in the unscreened limit the fifth force's perturbation to the energy levels depends linearly on the mass of the nucleus (as can be seen in Eq.~\eqref{energy-levels}), and a proton is heavier than a muon by a factor of 9.  In the limit in which the proton is unscreened, then, hydrogen bounds are superior, although this space is already excluded by other experiments.

As long as the Compton wavelength of the scalar field is much longer than the typical muon-electron separation (one Bohr radius $a_0$), we can use Eq.~\eqref{dE-unscreened} to find the general bound on the chameleon parameter
\begin{equation}
    M \gtrsim 550~\mathrm{GeV}~.
\end{equation}
At small values of $\Lambda$, however, the chameleon field becomes short-ranged inside the vacuum chamber, and the fifth force is exponentially suppressed by distance.  Precisely when this occurs depends on the experimental details of the muonium spectroscopy measurement~\cite{Meyer:1999cx}, which are summarised in Table~\ref{tab:params}.  The determination of $\phi_\mathrm{out}$ proceeds in the same way as the previous section, and the resulting bounds on chameleon parameter space are shown in Fig.~\ref{fig:chameleon-bounds}.

\section{Symmetron}

\begin{figure}[t]
    \centering
    \includegraphics[width=0.7\textwidth]{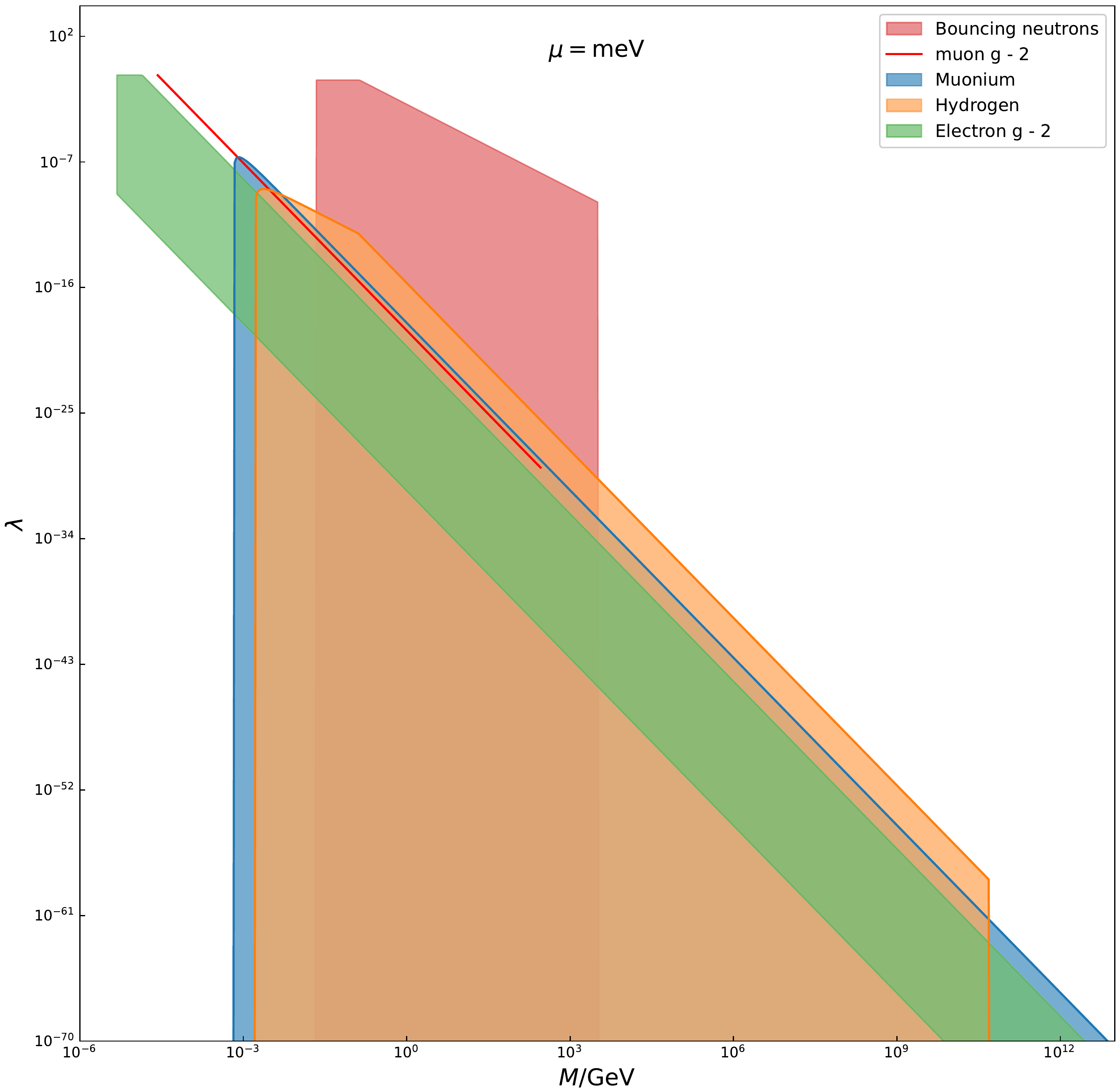}
    \caption{\scriptsize Bounds on symmetron models with mass at the dark energy scale $\mu = \mathrm{meV}$.  We see that hydrogen is the leading constraint when the proton is unscreened.  Meanwhile muonium bounds are comparable to those from electron $g-2$ experiments~\cite{Brax:2018zfb}.}
    \label{fig:symmetron-meV}
\end{figure}

In the past two sections we applied our results to a specific model, the chameleon, for concreteness.  In this section we compute bounds on another theory of modified gravity, the symmetron.  This model exhibits screening similar to, but distinct from, the chameleon.

The symmetron's couplings are given by its self-interaction potential $V(\phi)$ and matter coupling $A(\phi)$:
\begin{equation}
    V_\mathrm{symm}(\phi) = - \frac{1}{2} \mu^2 \phi^2 + \frac{1}{4} \lambda \phi^4~, \quad \quad A_\mathrm{symm}(\phi) = \frac{1}{2 M^2} \phi^2~.
\end{equation}
The symmetron's Lagrangian is symmetric under $\phi \to - \phi$.  However, examining the field's effective potential shows that this symmetry can be spontaneously broken in regions with sufficiently small matter density $\rho$: 
\begin{equation}
    V_\mathrm{eff}(\phi) = \frac{1}{2} \left( \frac{\rho}{M^2} - \mu^2 \right) \phi^2 + \frac{1}{4} \lambda \phi^4~.
\end{equation}
In regions where the ambient matter density is large, $\rho > \mu^2 M^2$, the sign of the quadratic term in the effective potential is positive and is minimised when $\phi = 0$.  However, in underdense regions, where $\rho < \mu^2 M^2$, the quadratic term becomes negative, giving an effective potential that spontaneously breaks the $\phi \to - \phi$ symmetry.  In this case, given sufficient room the field rolls to a vacuum expectation value (vev)
\begin{equation}
    v = \pm \frac{\mu}{\sqrt \lambda} \sqrt{1 - \frac{\rho}{\mu^2 M^2}}~.
\end{equation}
This is the ambient field value inside a vacuum chamber, provided that the chamber is larger than the symmetron's Compton wavelength $\approx \mu^{-1}$.  Scalar fluctuations about the vev have an effective mass
\begin{equation}
    m_\mathrm{eff} = \sqrt{2} \mu \sqrt{1 - \frac{\rho}{\mu^2 M^2}}~.
\end{equation}
Using $\phi_\mathrm{out} = v$ and this expression for $m_\mathrm{eff}$ in Eq.~\eqref{field-profile}, the symmetron screening factor for a uniform-density, spherical object of mass $m_\mathrm{N}$ and radius $R$ is
\begin{equation}
    \lambda_\mathrm{scr} = \min \left( \frac{4 \pi R M^2}{m_\mathrm{N}} , 1\right)~.
\end{equation}

\begin{figure}
    \centering
    \includegraphics[width=1.\textwidth]{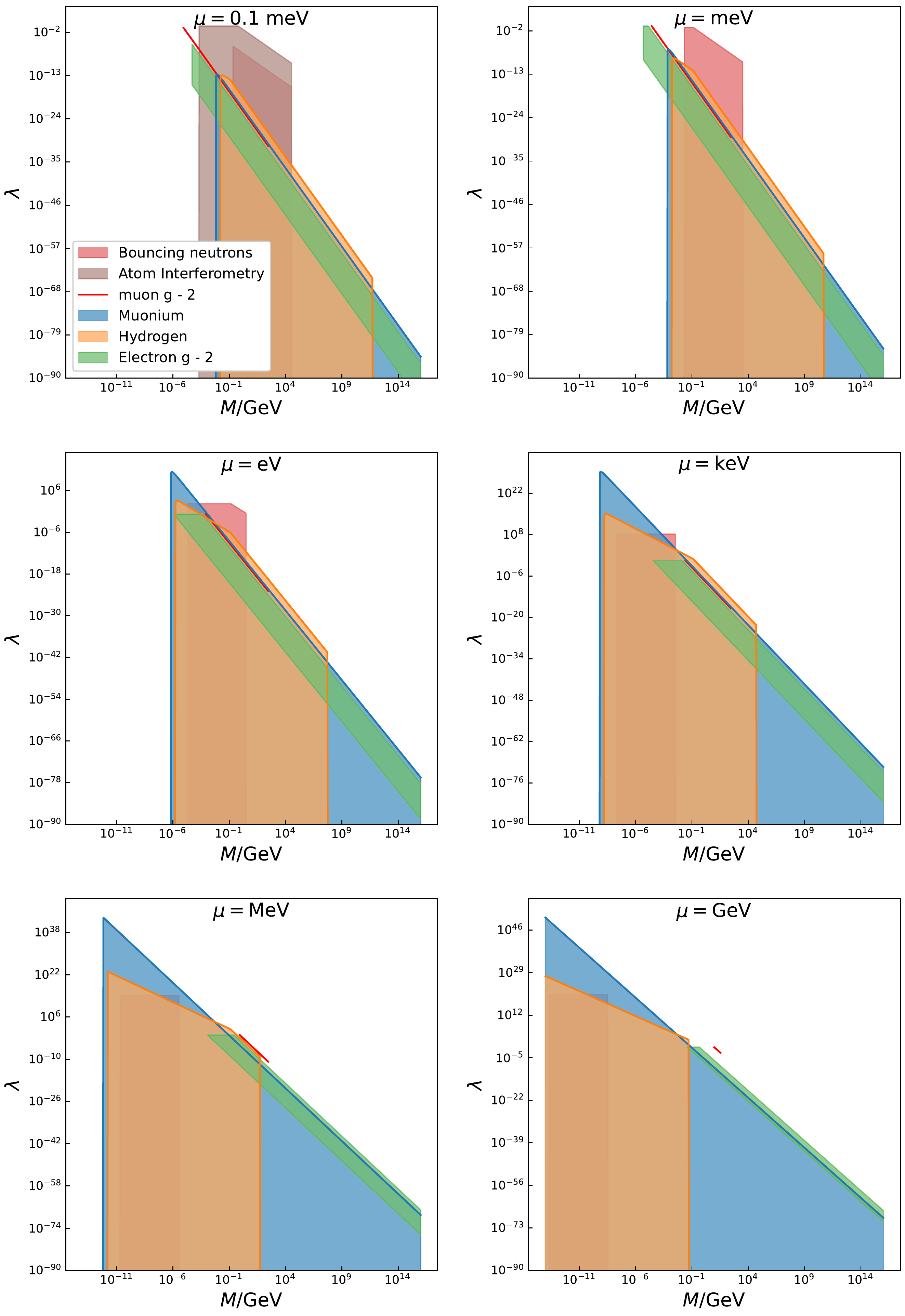}
    \caption{\scriptsize Bounds on symmetron parameters.  It can be seen that experiments employing extended objects test a limited range of symmetron masses, while those employing only fundamental particles access a much wider range of parameter space.}
    \label{fig:symmetron-all}
\end{figure}

If the vacuum chamber is smaller than the symmetron's Compton wavelength $\mu^{-1}$, the field remains at the false vacuum $\phi = 0$ everywhere~\cite{Upadhye:2012rc}.  This is because the stable field configuration is the one that minimises the energy, and therefore balances energy in the field's potential $V_\mathrm{eff}$ against gradient energy $(\vec \nabla \phi)^2$.  Achieving a non-zero field profile necessitates that the gradient energy cost of rolling to a non-zero value is outweighed by the energy savings of the field sitting at the true vacuum $\phi = v$ in the central regions of the vacuum chamber.  In a small vacuum chamber where $R_\mathrm{vac} < \mu^{-1}$, there is insufficient room for this to occur, and the field consequently remains at $\phi = 0$ everywhere inside the chamber.

With these considerations for the symmetron field, the computation of the constraints from hydrogen and muonium spectroscopy proceeds in exactly the same way as before.  One of the theoretically best-motivated values of $\mu$ is to set it to the dark energy scale, $\mu = \mathrm{meV}$, as the vacuum fluctuations of the symmetron would account for dark energy.  The bounds we obtain in this case are plotted in Fig.~\ref{fig:symmetron-meV}.  We see that the bounds from muonium
%are competitive with those from colliders, and 
are slightly superior to bounds from electron $g-2$ experiments.  Meanwhile, we see that hydrogen, when the proton is unscreened (that is, for $M \gtrsim 100 ~\mathrm{MeV}$) is the leading experimental constraint.

In Fig.~\ref{fig:symmetron-all} we show bounds for a range of $\mu$ values spanning 14 orders of magnitude.  It is remarkable that a single experiment can constrain such a wide range in $\mu$.  This is possible for the following reason.  A laboratory experiment employing extended objects can only constrain a narrow range $L_\mathrm{system} < \mu^{-1} < R_\mathrm{vac}$, where $L_\mathrm{system}$ is the size of the system being measured and $R_\mathrm{vac}$ is the size of the chamber the experiment is performed in.  This is because a Compton wavelength $\mu^{-1}$ that is smaller than the distance between an experiment's source and test masses leads to an exponentially suppressed symmetron force as $e^{-\mu r}$.  On the other hand, we have already seen that a Compton wavelength $\mu^{-1}$ that is larger than the vacuum chamber leads to a vanishing vev and matter coupling.  For an atom interferometry test, this hierarchy of scales is of order $L_\mathrm{system} / R_\mathrm{vac} \approx \frac{\mathrm{mm}}{10~ \mathrm{cm}} \approx 10^{-2}$.  For a hydrogen or muonium spectroscopy experiment, this hierarchy is much larger $L_\mathrm{system} / R_\mathrm{vac} \approx \frac{\text{\AA}}{10~ \mathrm{cm}} \approx 10^{-9}$, allowing sensitivity to symmetron models with a much wider range of $\mu$ values.  This trait is shared by other experiments that probe very short length scales, notably bouncing neutrons~\cite{Jenke:2014yel,Jenke:2020obe} and $g-2$ measurements~\cite{Brax:2018zfb}.

In the case of muonium, the constraint takes a particularly simple form as there is no screening of the nucleus to account for.  In this case, the perturbation to the $1s-2s$ energy gap is
\begin{equation}
    \delta E_\mathrm{1s-2s} = \frac{3 \mu^2}{\lambda M^2} \frac{m_\mu m_e}{16 \pi a_0}~,
\end{equation}
so long as the vacuum's density is below the symmetron's critical density, $\rho_\mathrm{vac} < \mu^2 M^2$, and the symmetron Compton wavelength lies between the average muon - electron separation distance and the vacuum chamber size,
$1~\mathrm{\AA} < \mu^{-1} < 1~\mathrm{cm}$.
The constraint may then be expressed economically as
\begin{equation}
    \frac{\mu}{\sqrt{\lambda} M} < 10^{-12}~\mathrm{eV}~,
\end{equation}
so long as the above-mentioned inequalities are satisfied.
This bound can be observed as a diagonal line in constraint plots for a range of symmetron masses in Figs.~\ref{fig:symmetron-meV} and \ref{fig:symmetron-all}.

We observe that the muonium bounds are competitive with the electron $g - 2$ experiment and (along with the $g - 2$ experiment) are the leading source of constraints when the symmetron mass is greater than an MeV.

\section{Conclusions}
In this paper we have computed the perturbations to the spectra of hydrogen-like systems due to screened scalar fields.  We applied our results first to hydrogen, as its $1s-2s$ energy level spacing is known to great precision.  We then applied the same techniques to muonium.  Although muonium's spectrum has not been measured to the same degree of accuracy, the fact that it is composed entirely of fundamental particles makes it an excellent probe of screened theories.  We have presented bounds for two prototypical examples of screened theories: the chameleon and the symmetron.

We have shown that the muonium bounds on these theories are approximately an order of magnitude stronger than those derived electron g-2 experiments, as long as the scalar field's Compton wavelength is larger than the average muon-electron separation distance in muonium.  Furthermore, the new muonium bounds rule out about half of the viable chameleon models that eliminate the muon $g - 2$ tension~\cite{Brax:2018zfb}, although it is worth noting that those models are already in some tension with bounds from kaon decays.  That being said, those bounds originate from particle physics experiments where the validity of the screened modified gravity models as effective field theories is not guaranteed. Here we have focused on low energy physics only and therefore can trust the effective description.

This result makes it clear that a reduction in the uncertainty of muonium's $1s-2s$ transition energy by one to two orders of magnitude could play a major role in constraining, or ruling out, relevant screened modified gravity theories. This would directly test the models that seek to explain the muon anomalous magnetic moment results.  It would also test chameleon models that are currently being investigated within the context of cosmic voids~\cite{Tamosiunas:2022tic}.

In the future, it would also be of interest to include couplings to other Standard Model gauge fields, most notably the photon.  This would modify the scalar field profile that is sourced by the nucleus and would enable spectroscopy measurements to also place bounds the photon-scalar coupling.  This is left for future work.

{\small {\bf Acknowledgments} We are grateful for helpful discussions with Clare Burrage, Joerg Jaeckel, and Jeremy Sakstein.}

\renewcommand{\em}{}
\addcontentsline{toc}{section}{References}
\bibliography{main}
% need to \input the bbl file for PRD to work

\end{document}